\documentclass{webofc}
\usepackage[varg]{txfonts}   
\def\mbf#1{\mbox{\boldmath$#1$}}
\def\mbfs#1{\mbox{\scriptsize\boldmath$#1$}}
\def\vh{\mbf{h}}
\def\vr{\mbf{r}}

\begin{document}
\title{Numerical Precision Effects on GPU Simulation of Massive Spatial Data, Based on the Modified Planar Rotator Model}
%
%

\author{\firstname{Mat\'u\v{s}} \lastname{Lach}\inst{1}\fnsep\thanks{\email{matus.lach@student.upjs.sk}} \and
        \firstname{Michal} \lastname{Borovsk\'y}\inst{1} \and
        \firstname{Milan} \lastname{\v{Z}ukovi\v{c}}\inst{1}
}

\institute{Department of Theoretical Physics and Astrophysics, P. J. \v{S}af\'arik University,\\ \phantom{$^1$}Park Angelinum 9, 041 54 Ko\v{s}ice, Slovakia}

\abstract{%
  The present research builds on a recently proposed spatial prediction method for discretized two-dimensional data, based on a suitably modified planar rotator (MPR) spin model from statistical physics. This approach maps the measured data onto interacting spins and, exploiting spatial correlations between them, which are similar to those present in geostatistical data, predicts the data at unmeasured locations. Due to the short-range nature of the spin pair interactions in the MPR model, parallel implementation of the prediction algorithm on graphical processing units (GPUs) is a natural way of increasing its efficiency. In this work we study the effects of reduced computing precision as well as GPU-based hardware intrinsic functions on the speedup and accuracy of the MPR-based prediction and explore which aspects of the simulation can potentially benefit the most from the reduced precision. It is found that, particularly for massive data sets, a thoughtful precision setting of the GPU implementation can significantly increase the computational efficiency, while incurring little to no degradation in the prediction accuracy.
}
\maketitle
\section{Introduction}
\label{intro}
Enormous amounts of Earth observation data collected by remote sensing techniques, such as geographical, natural resources, land use, and environmental, require highly efficient (preferably real-time) processing \cite{RSE1}. Such processing often involves the reconstruction of data which is missing due to various reasons, e.g. equipment malfunctions or gaps in the coverage of the targeted area that appear as a result of restricted satellite paths or
bad weather conditions \cite{Atmos}. These massive data sets, however, cannot be efficiently
handled by the standard geostatistical methods, such as kriging \cite{Kriging}. The main drawbacks of such methods are the high computational complexity, difficulties to automatize the algorithms to work
without subjective user inputs and, in the case of non-Gaussian data, preprocessing requirements \cite{MBG}.

An alternative approach inspired by spin models in statistical physics has been reported in \cite{MPR}, based on the so-called modified planar rotator (MPR) model. This approach maps the measured data onto spins situated on a grid and exploits spatial correlations between them, which turned out to be similar to those present in the geostatistical data, to predict the missing data points. The prediction procedure then consists of conditional Monte Carlo simulations with a single parameter to be inferred, called reduced temperature. The MPR method has been shown to be particularly competitive and effective for non-Gaussian data \cite{MPR}. 

Furthermore, considering the fact that the spins in the MPR model interact only with their nearest neighbors, this method  is suitable for parallelization and can be implemented on graphical processing units (GPUs), which posses by far superior theoretical peak performance due to their inherently parallel architecture. In our recent work \cite{MPRGPU}, almost $500$ times speedup compared to the CPU implementation has been achieved. 

\section{Model and Methods}
\label{sec-1}
The Hamiltonian of the standard planar rotator is described by
\begin{equation}
\label{eq_1}
\mathcal{H} = -J\sum_{\langle i,j\rangle} \textbf{S}_i \cdot \textbf{S}_j = -J\sum_{\langle i,j\rangle} \cos (\phi_i - \phi_j),
\end{equation} 
where $\textbf{S}_i, \textbf{S}_j$ are neighboring spins at the sites $i$ and $j$ represented by unit vectors in the $XY$ plane, $\phi_{i}, \phi_{j} \in [0, 2\pi]$ are their turn angles and $J$ is the exchange interaction parameter. It can be modified to account for the spatial correlations typical for geostatistical data, simply by introducing the parameter $q \in (0, {1}/{2}]$ into the Hamiltonian, such that it becomes
\begin{equation}
\label{eq_2}
\mathcal{H} = -J\sum_{\langle i,j\rangle} \cos [q(\phi_i - \phi_j)].
\end{equation} 
It has been shown \cite{JOPCS}, that a spin system described by this modified Hamiltonian displays temperature-controlled short-range correlations, which are typical for the geostatistical data. The temperature is the only simulation parameter and needs to be inferred from the incomplete data. Then, a conditional Monte Carlo simulation, in which only the spins at the prediction locations are allowed to change their states, is performed using the so-called hybrid algorithm \cite{MPR}. By averaging over multiple configurations in equilibrium, one can obtain predictions on the discretized two-dimensional missing data.
 
The simulation codes in this study were written in the C++ language and parallelized on NVIDIA GPU GTX 980 using the Compute Unified Device Architecture (CUDA) framework. Details of the GPU implementation can be found in \cite{MPRGPU}. A large part of the achieved speedup in \cite{MPRGPU} is due to the fact that the lowering of the arithmetic precision has far greater potential impact on the GPUs, compared to the CPUs. Below, we look into the effects of different precision setups on the efficiency and accuracy of the computations. 

There are two main areas which can potentially benefit from the lowering of the arithmetic precision. The spin variables and the system specific energy, defined as $e = \langle\mathcal{H} \rangle/{L^2}$,
where $L$ is the linear size of the square grid, can both be represented as either 64-bit double precision (DP) or 32-bit single precision (SP) variables. Since the SP variables require only half of the memory needed by DP variables, the speed of all read and write memory operations can be improved considerably. The second area is the precision of the functions used during the course of simulations. In C++, it is possible to use precision-specific versions of common functions, such as trigonometric, exponential etc. SP functions sacrifice accuracy for decreased computation time. In addition, CUDA offers so-called hardware-intrinsic versions of these functions, which utilize mapping to fewer hardware instructions, increasing speed at the cost of accuracy and different special case handling. These hardware intrinsic functions were used in all cases of the SP computations of the  functions. In the following we consider eight possible precision settings of the spin variables ($X_S$), the energy ($X_E$) and the functions ($X_F$), where $X_i = D$ or $S$, i.e., ($X_SX_EX_F$) ranging from full DP (DDD), through mixed precision to full SP calculations (SSS). 

\section{Results and Conclusions}
\label{sec-2}
The simulations were performed on synthetic Gaussian data $Z \sim N (m = 50, \sigma = 10)$ with exponential covariance given by
\begin{equation}
G_Z(\Vert \vh \Vert) = \sigma^2\exp(-\kappa \Vert \vh \Vert).
\end{equation}
Here $\Vert \vh \Vert$ is the Euclidean two-point distance, $\sigma^2$ is the variance, $\kappa$ is the inverse autocorrelation length (set to $\kappa = 0.2$) on square grids with the side length $L = 2^k, k = 5, 6, ..., 12$. The exponential covariance model is suitable for modeling rough spatial processes such as soil data \cite{WM}. 90\,\% of the original data were removed at random points and stored for later comparison to the predicted values.
\begin{figure*}
\centering
\hspace*{-1.5cm}
\includegraphics[width=16.5cm,clip]{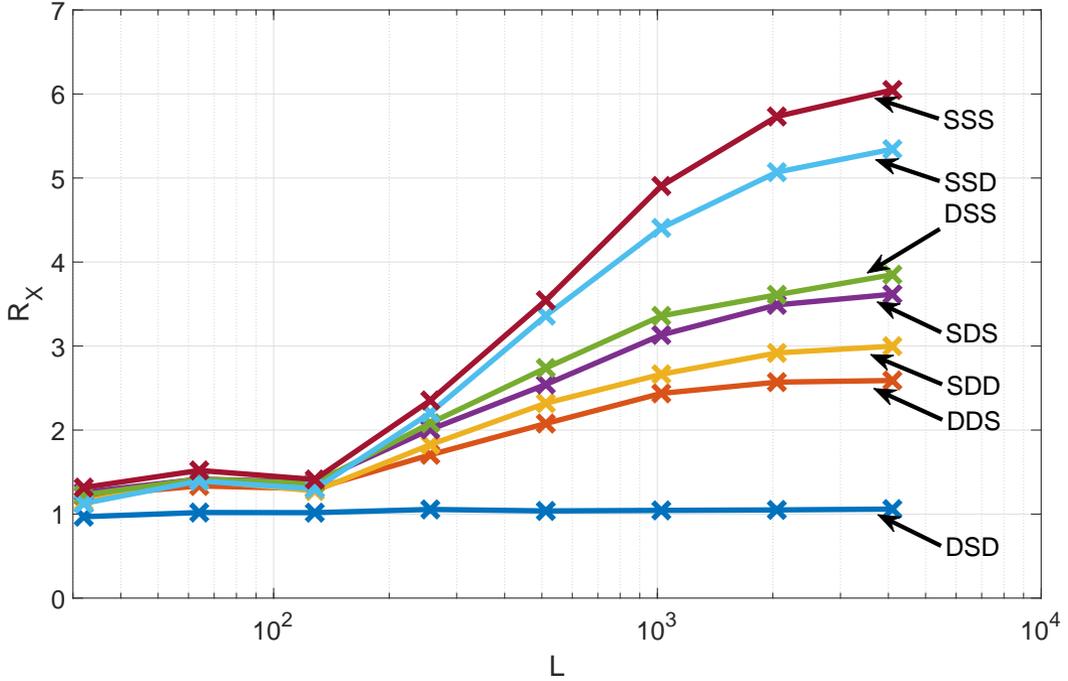}
\caption{Relative speedups $R_x$ of various degrees of mixed precision $X$ and full single precision (SSS) compared to full double precision (DDD) for various linear grid sizes from $L = 36$ to $L = 4096$.}
\label{fig-speedup}       
\end{figure*}
Figure~\ref{fig-speedup} shows that lowering the arithmetic precision has a pronounced effect on the speed of the simulations. We record a relative speedup $R_X$ defined as the ratio of the GPU-time required by the precision setting $X$ and DDD, i.e., $R_X = {t_{DDD}^{GPU}} / {t_X^{GPU}}$. The speedup increases as larger portions of the simulations are ported to SP and it is further amplified by the increasing number of processed data points, reaching 6x speedup compared to DDD for the largest considered grid size of $L = 4096$. 
\begin{figure*}
\centering
\sidecaption
\includegraphics[width=9cm,clip]{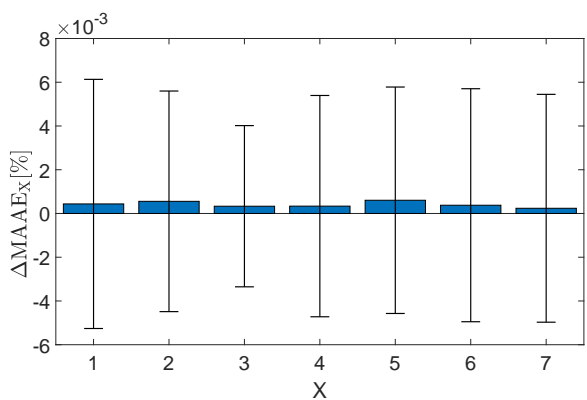}

\caption{Relative increase of the mean average absolute error compared to full double precision for all the possible configuration of mixed precision and the full single precision X: 1 -- DSD, 2 -- DDS, 3 -- SDD,\qquad  4 -- SDS,  5 -- DSS,  6 -- SSD, 7 -- SSS}
\label{fig-errors}       
\end{figure*}

To evaluate the accuracy of the MPR-based prediction, we compare the predicted values to the true ones and compute the average absolute error (AAE) defined as
\begin{equation}
\mathrm{AAE} = \frac{1}{P}  \sum_{\mbfs{r}_p \in G_p} |\epsilon(\vr_p)|,
\end{equation}
where $\epsilon(\vr_p) = Z(\vr_p) - \hat{Z}(\vr_p)$ is the difference between the true value $Z(\vr_p)$ and the predicted value $\hat{Z}(\vr_p)$ at the site $\vr_p$ and $P = 0.9L^2$ is the number of prediction sites. For each complete data set we generate $M=100$ different sample configurations with missing data and calculate the mean AAE (MAAE). To evaluate the variation of the prediction error with the lowering of numerical precision, we compute the relative change of MAAE compared to its value obtained from full DP simulations in $\%$, as $\Delta \mathrm{MAAE}_{\mathrm{X}} = (\mathrm{MAAE}_{\mathrm{X}} / \mathrm{MAAE}_{\mathrm{DDD}} - 1)100\%$, where $\mathrm{X}=1,\ldots,7$, and plot the results in Figure~\ref{fig-errors}. The bars are sorted by the increasing speedup from Figure~\ref{fig-speedup} with the slowest on the left and the fastest on the right. We note that the increase of the prediction error is only of the order of $10^{-4}$\,\% with relatively large error bars, which indicate that all the values of $\Delta MAAE$ coincide within the error bars and thus there is no trend of increasing the error with lowering precision. 

Thus, we conclude there is no practical reason to use any degree of mixed precision or full DP, since full SP with hardware intrinsic functions provides by far superior simulation performance with negligible error increase. We intend to extend this assessment by further lowering of the arithmetic precision to the 16-bit half precision variables and functions, which are available on the latest NVIDIA GPUs,  in expectation of further considerable increase of the computational speedup over the CPU implementation.

\subsection*{Acknowledgement}
This work was supported by the Scientific Grant Agency of Ministry of Education of Slovak Republic (Grant No. 1/0531/19).

%

\begin{thebibliography}{8}
%
\bibitem{RSE1} R. E. Rossi, J. L. Dungan, L. R. Beck, Remote Sensing of Environment, \textbf{49}, 32--40 (1994)
\bibitem{Atmos} M. Jun, M. L. Stein, Atmospheric Environment, \textbf{38}, 4427--4436 (2004)
\bibitem{Kriging} H. Wackernagel, \textit{Multivariate Geostatistics}, (Springer, New York, 2003) 79--100
\bibitem{MBG} P. J. Diggle, P. J. Ribeiro Jr.,  \textit{Model-based Geostatistics. Series: Springer series in statistics}, (Springer, New York, 2007) 199--212
\bibitem{MPR} M. Žukovič, D. T. Hristopulos, Physical Review E \textbf{98}, 062135-1--23 (2018)
\bibitem{MPRGPU} M. Žukovič, M.  Borovský, M. Lach, D. T. Hristopulos, Math Geosci (2019), https://doi.org/10.1007/s11004-019-09835-3
\bibitem{JOPCS} M. Žukovič, D. T. Hristopulos, Journal of Physics: Conference Series \textbf{633}, 012105 (2015)
\bibitem{WM} B. Minasny, A. B. McBratney, Geoderma \textbf{128}, 192--207 (2005) 
%
\end{thebibliography}
\end{document}